\RequirePackage[2020-02-02]{latexrelease}
\documentclass[twocolumn,prl,amsmath,amssymb,showpacs,superscriptaddress,floatfix]{revtex4}
\usepackage{graphicx}
\usepackage{bm}
\usepackage{lineno,hyperref}
\usepackage{epsf}
\usepackage{xcolor}

\begin{document}
\title{Ultrasonic study and molecular simulation of propylene glycol at pressure up to 1.4 GPa}

\author{Yu. D. Fomin \footnote{Corresponding author: fomin314@mail.ru}}
\affiliation{Vereshchagin Institute of High Pressure Physics,
Russian Academy of Sciences, Kaluzhskoe shosse, 14, Troitsk,
Moscow, 108840, Russia } 

\author{I. V. Danilov}
\affiliation{Vereshchagin Institute of High Pressure Physics,
Russian Academy of Sciences, Kaluzhskoe shosse, 14, Troitsk,
Moscow, 108840, Russia } 

\author{E. L. Gromnitskaya}
\affiliation{Vereshchagin Institute of High Pressure Physics,
Russian Academy of Sciences, Kaluzhskoe shosse, 14, Troitsk,
Moscow, 108840, Russia }

\date{\today}

\begin{abstract}

We report an ulsrasonic measurements of density and bulk modulus of propylene glycol at room temperature and
at the temperature of liquid nitrogen combined with molecular dynamics simulations with two different force fields.
We find that experimental density of propylene glycol at room temperature is well described within COMPASS force fields 
simulations, while the bulk modulus from simulation deviates from the experimental one. Number of hydrogen bonds
in propylene glycol is also evaluated.

\end{abstract}


\maketitle

\section{Introduction}

Propylene glycol (PG) is an important liquid for many technological applications, including food industry, resin production, pharmacy, etc. It is also widely used as antifreeze. 
Al of this makes understanding of the properties of PG to be of a great importance. And indeed there is a plenty of publications devoted to different
properties of PG. At the same time there are still a lot of properties of PG which are still badly understood. 

Propylene glycol has a chemical formula $C_3H_8O_2$. It is a molecular liquid, which can be easily transformed into glass (the glass transition temperature at ambient pressure
is $T_g=165$ K). A molecule of PG contains two hydroxyl groups, which makes it possible to form hydrogen bonds (H-bonds) in PG. Although H-bonds
are weaker than other types of bonding they can strongly affect the properties of a liquid \cite{h1,h2,h3,h4}. H-bonds can affect the orientational correlations
between molecules of a liquid and glasses. The relaxational dynamics of H-bonds strongly affects the dynamical properties of liquids. H-bonds also
can be responsible for the properties of a liquid as a solvent, as it is assumed in the case of water \cite{h5}. Apparently, a lot of attention of researchers was
paid to investigation of H-bonding in different liquids, both experimentally and theoretically. For instance, in Refs. \cite{h6,h7,h8,h9,h10,h11,h12,h13,h14} experimental
study of H-bonds in liquid alcohols was reported. These works were complemented by molecular dynamics simulation \cite{h15,h16,h17}. However, the problem of
the behavior of H-bonds still remains unclear, since different works give different, sometimes opposite results (see, for instance, Refs. \cite{met1,met2,met3,met4,met5} for 
hydrogen bonds in methanol).

In the present work we perform ulstrasonic measurements and molecular simulations of PG at room temperature and at the temperature of liquid nitrogen up
to the pressure of $1.4$ GPa. We measure the equation of state, compressibility and its baric derivative and compare the results to the ones from computer
simulation. Molecular simulation also allows us to monitor microscopic structure of PG and evaluate the number of H-bonds in a wide range of pressures.

\section{System and Methods}

\subsection{Experimental methods}

The experimental part of the work was carried out using a high-pressure ultrasonic piezometer. It was created on the basis of a press and a piston-cylinder high-pressure chamber, placed in a thermostat for low-temperature measurements \cite{i1,i2}. This setup makes it possible to carry out experiments up to pressures of 2 GPa and in the temperature range of 77–310 K both in the isothermal compression and isobaric heating regimes. In this work, we performed isothermal compression up to 1 GPa at constant temperatures of 77 K (liquid nitrogen temperature) and 295 K (room temperature). The pressure measurement error was no more than 0.05 GPa, that is, $5 \%$, and the temperature was controlled with an accuracy of 1 K. To measure the temperature, we used 4 copper-constantan thermocouples: 2 were glued in the immediate vicinity of the sample and 2 more were glued to punches that pressed on sample, to control the temperature gradient. Since propylene glycol is a liquid under normal conditions, we used capsules to contain the substance. The capsule is a thin-walled Teflon cylinder with an outer diameter of 18 mm and an inner diameter of 16 mm, which is closed on both sides with copper caps 1 mm thick and a rubber ring as a seal. 
Propylene glycol $99.5 \% $ pure was purchased from Sigma-Aldrich. The capsule with the substance was placed in a high-pressure chamber and compressed from both sides by punches, on the ends of which piezoplates of lithium niobate $LiNbO_3$ were glued. They were both generators and receivers of the ultrasonic signal. For longitudinal and transverse ultrasonic waves, different plates were used, with different thicknesses and different resonant frequencies: 5 MHz for transverse and 10 MHz for longitudinal ultrasonic waves. The transit time of ultrasound through the sample was measured with an accuracy of 1 ns. The length of the sample was also measured - for this we used 2 micrometers with a measurement accuracy of 5 $\mu m$. From the length and transit time, the ultrasound velocity was obtained, then, in the approximation of a homogeneous isotropic medium \cite{i3,i4} the adiabatic bulk modulus $B_s$ was calculated:

\begin{equation} \label{eq1}
B_s=\rho v_l^2- \frac{4}{3} \rho v_t^2,
\end{equation}
where $\rho$ is the density of a substance, $v_l$ is longitudinal ultrasonic wave velocity and $v_t$ is the transverse ultrasonic wave velocity.

In Supplementary materials we describe some details of the experimental derivation of density of PG.

\subsection{Molecular simulation}

\begin{figure}

\includegraphics[width=6cm, height=6cm]{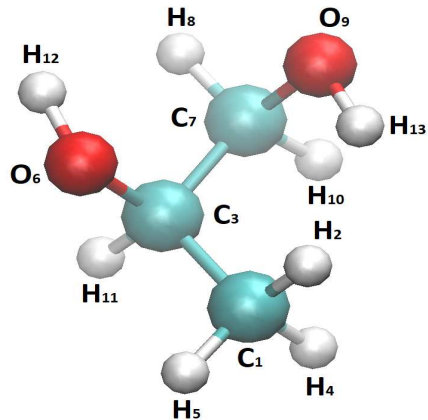}%

\caption{\label{pg-mol} An image of a molecule of PG with notation of atoms.}
\end{figure}

In the computational part of the work we simulated a system of 1000 molecules of PG in a cubic box with periodic boundary conditions. 
Two different force fields (FF) were used: COMPASS \cite{compass} and Charmm \cite{charmm}. Figure \ref{pg-mol} shows a molecule
of PG with notation of all atoms.

In both FFs the system was equilibrated for $1 \cdot 10^7$ steps with the time step $dt=0.1$ fs. Then the system was simulated
for $2 \cdot 10^7$ step for calculation of its properties. The bonds involving hydrogens were constrained via Shake algorithm \cite{shake}. 
Firstly the system was simulated at temperature $T=295$ K at several densities in canonical ensemble (constant number 
of particles N, volume V and temperature T). Equation of state was calculated and compared to the experimental data. 
This equation of state was approximated by polynomial functions to calculate the isothermal bulk modulus $B_T=\rho \left( \frac{dP}{d \rho} \right)_T$ and
the isothermal baric derivative $\left( \frac{dB}{dP} \right)_T$. The details of approximation are given in Supplementary materials.

When the simulation at $T=295$ K was finished the system was quenched to a low temperature $T=77$ K. The cooling was made with within
$5 \cdot 10^7$ steps, i.e. with the cooling rate $4.36 \cdot 10^9$ $K/s$. Equation of state (EoS), bulk modulus and
baric derivative were calculated at the low temperature too.

\begin{figure}

\includegraphics[width=4cm, height=4cm]{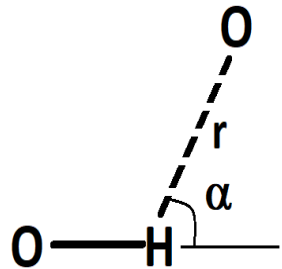}%

\caption{\label{hbond} The definition of hydrogen bonds in the system.}
\end{figure}

Figure \ref{hbond} shows a definition of hydrogen bonds in the system. We assume that a hydrogen bond is formed if two conditions
are fulfilled: i) the distance $r<3.5$ $\AA$ and (ii) the angle $\alpha <30^o$. 

The structure of the system was characterized by center of mass radial distribution functions (RDFs) and partial radial distribtuion functions of
different species (see Fig. \ref{pg-mol}). 

Hydrogen bonding in PG is discussed. The presence of H-bonds was analyzed based on geometrical criterion. Several definitions of H-bonds
in water were discussed in Ref. \cite{hbond-water}. In the present paper we adopted the $r - \alpha$ definition from this paper, i.e.
we assume that an H-bond is formed if the distance $r$ is below some cut-off value and the angle $\alpha$ is below some cut-off angle
(see Fig. \ref{hbond}). Following Ref. \cite{pg-ff} we choose $r_{cut}=3.5  \AA$ and $\alpha_{cut}=30^o$.

All simulations were performed using the LAMMPS simulation package
\cite{lammps}.

\section{Results and Discussion}

\begin{figure}

\includegraphics[width=8cm, height=6cm]{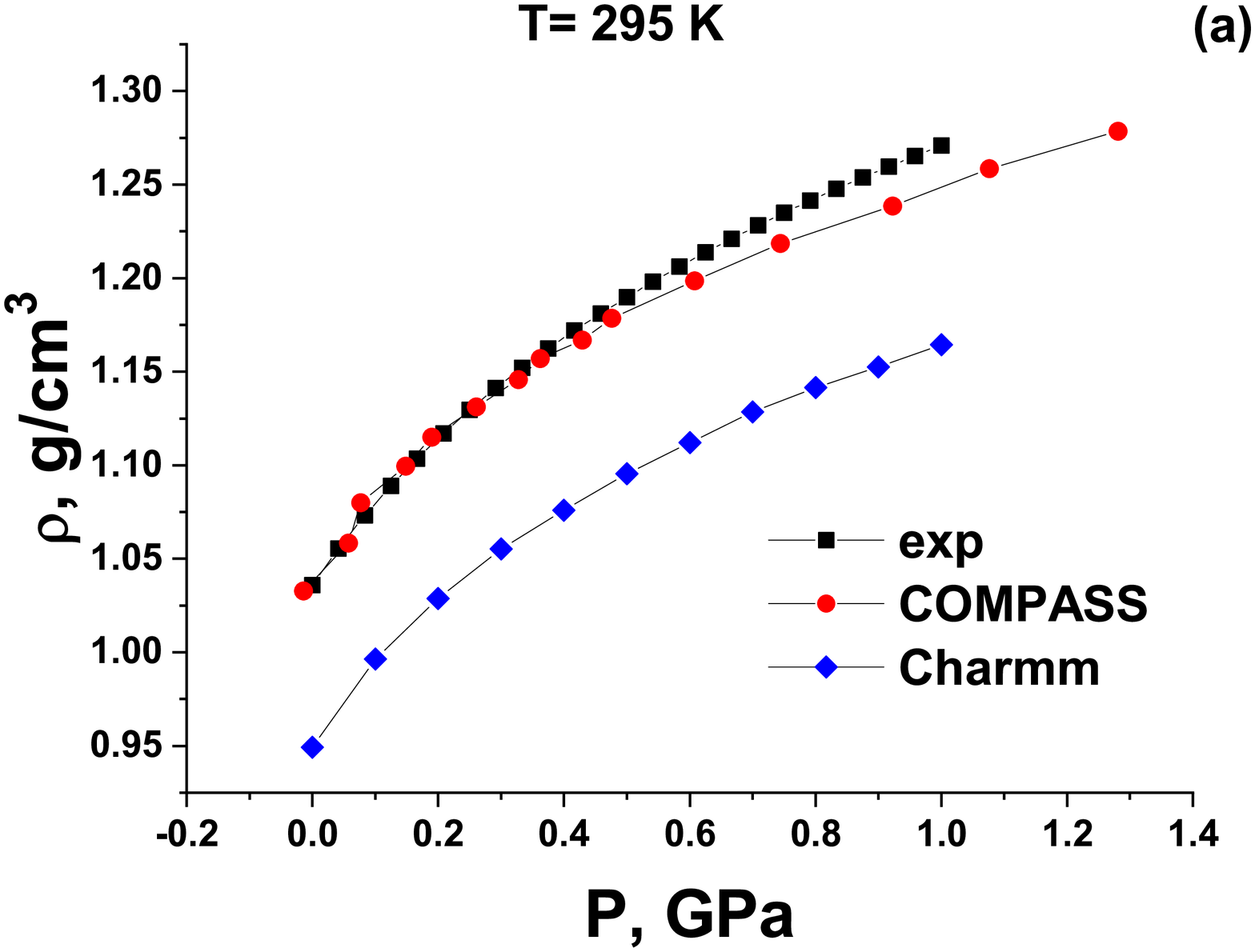}%

\includegraphics[width=8cm, height=6cm]{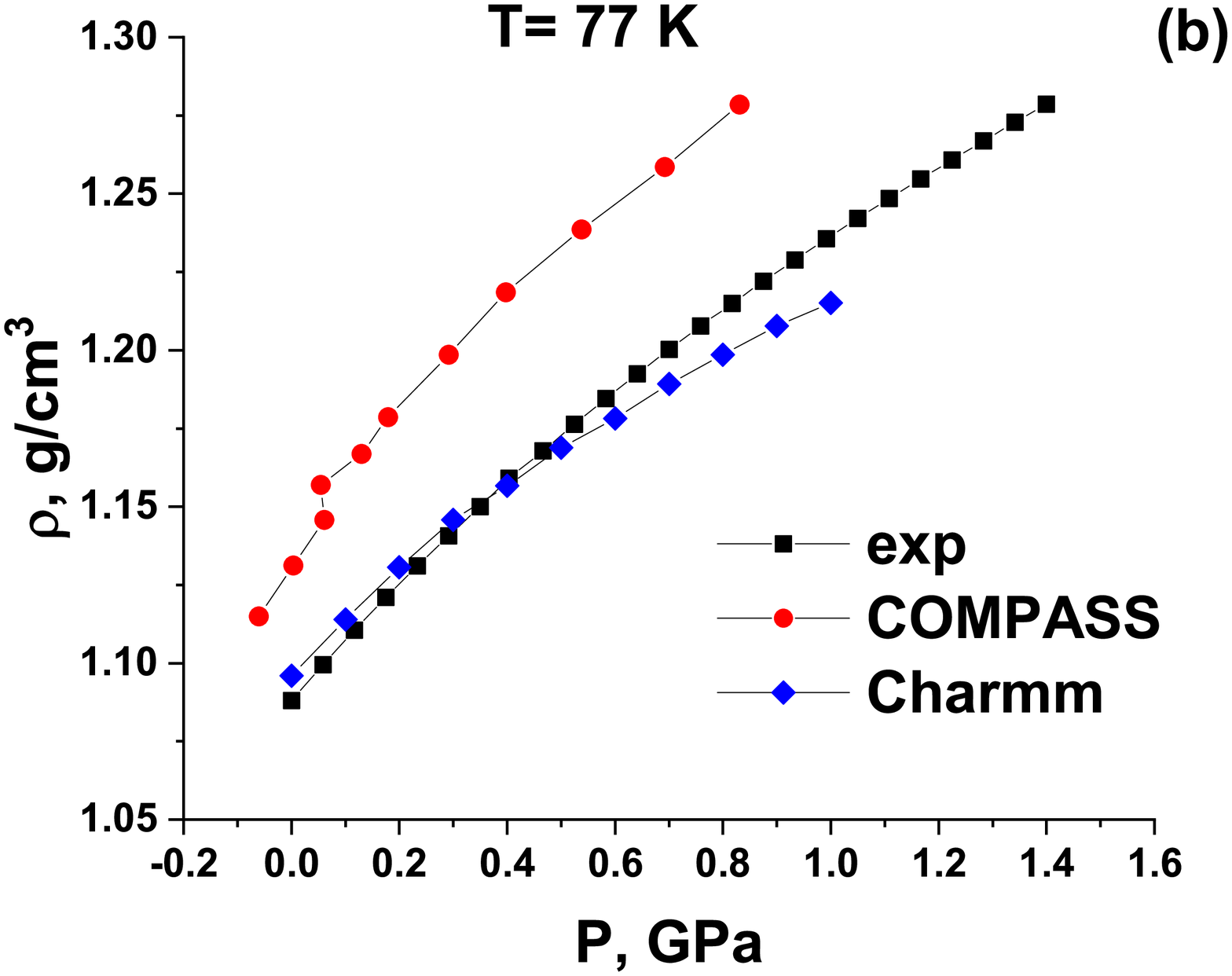}%

\caption{\label{eos-md-exp} Comparison of equation of state from experiments and molecular simulations with COMPASS and Charmm models at (a) $T=295$ K and (b) T=77 K.}
\end{figure}

Figure \ref{eos-md-exp} shows a comparison of experimental data with molecular simulation with two different force fields - COMPASS and Charmm. One can
see that at room temperature COMPASS force field perfectly coincide with experimental data up to the pressure of 0.5 GPa and stays within the error bars of 
experiment up to about 0.8 GPa. At the same time one can see that starting from $P \approx 0.5$ GPa the experimental density is higher than the 
one from simulation. Several reasons can induce this disagreement. In particular, the ultrasonic measurements of PG were carried out both on compression and
decompression modes. The intrinsic friction of the installation leads to overestimation of the pressure in the compression cycle and underestimation in the 
decompression mode. As a result we observe a hysteresis loop which is eliminated by averaging of the results. However, this method works
accurately in the pressure interval of $0.2-0.7$ GPa only. Errors can appear due to the uniaxial compression in low pressure regime and due
to very strong friction in the high pressure one. Thus, the discrepancy between the theoretical and experimental density at high pressures can 
be explained by the increased experimental error in pressure measurements. Basing on this reasoning we may suppose that experimental densities
are slightly overestimated at high pressures, and the ones from COMPASS FF modelling are closer to the real ones. Unfortunately, we are not aware of any measurements of
EoS of PG at elevated pressures by another experimental method and therefore we cannot verify our assumption by direct comparison of the data. 

The results of molecular simulations with Charmm FF strongly underestimate the density at $T=295$ K. The disagreement is about $10 \%$, which we consider
as unsatisfactory result.

\begin{figure}

\includegraphics[width=8cm, height=6cm]{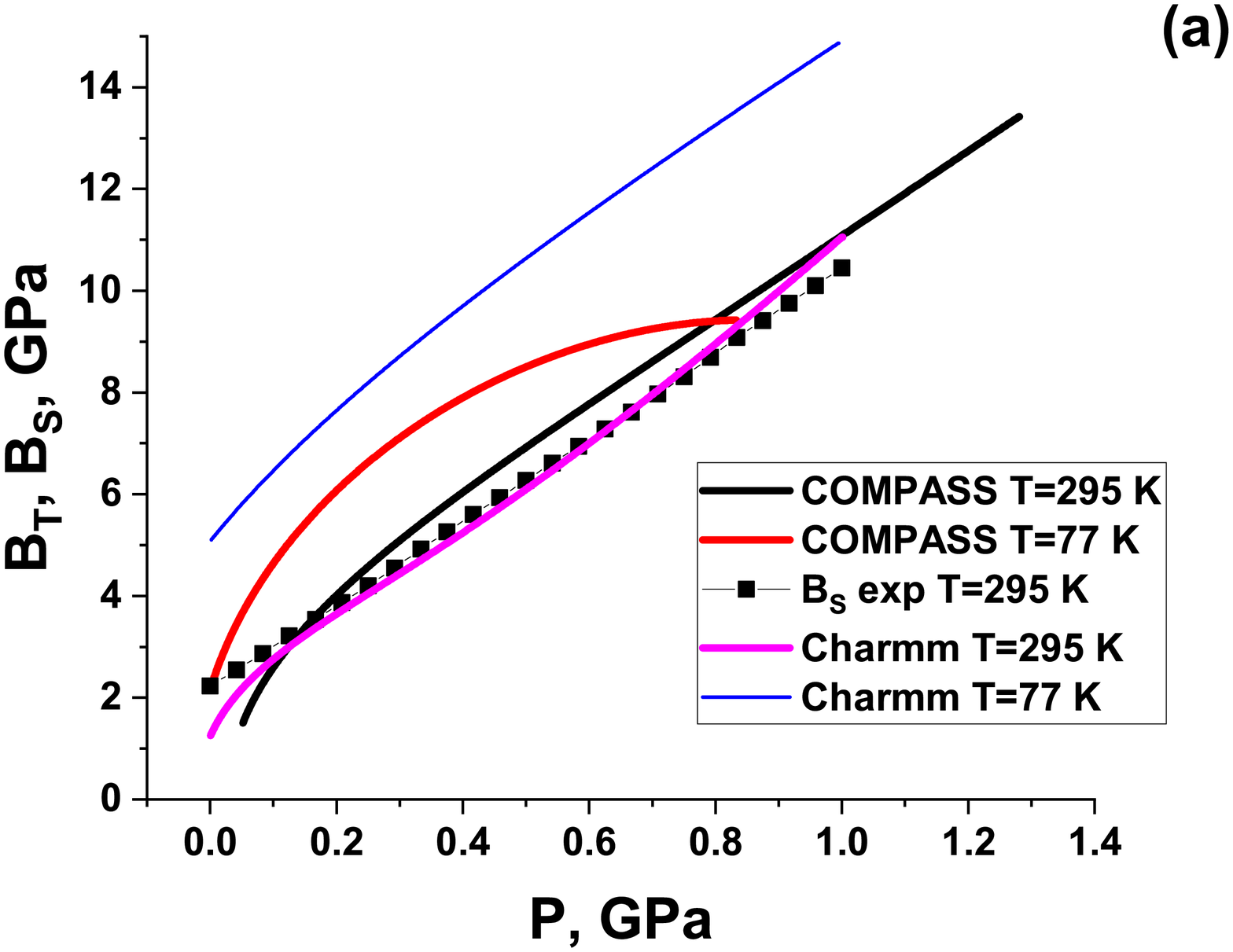}%

\includegraphics[width=8cm, height=6cm]{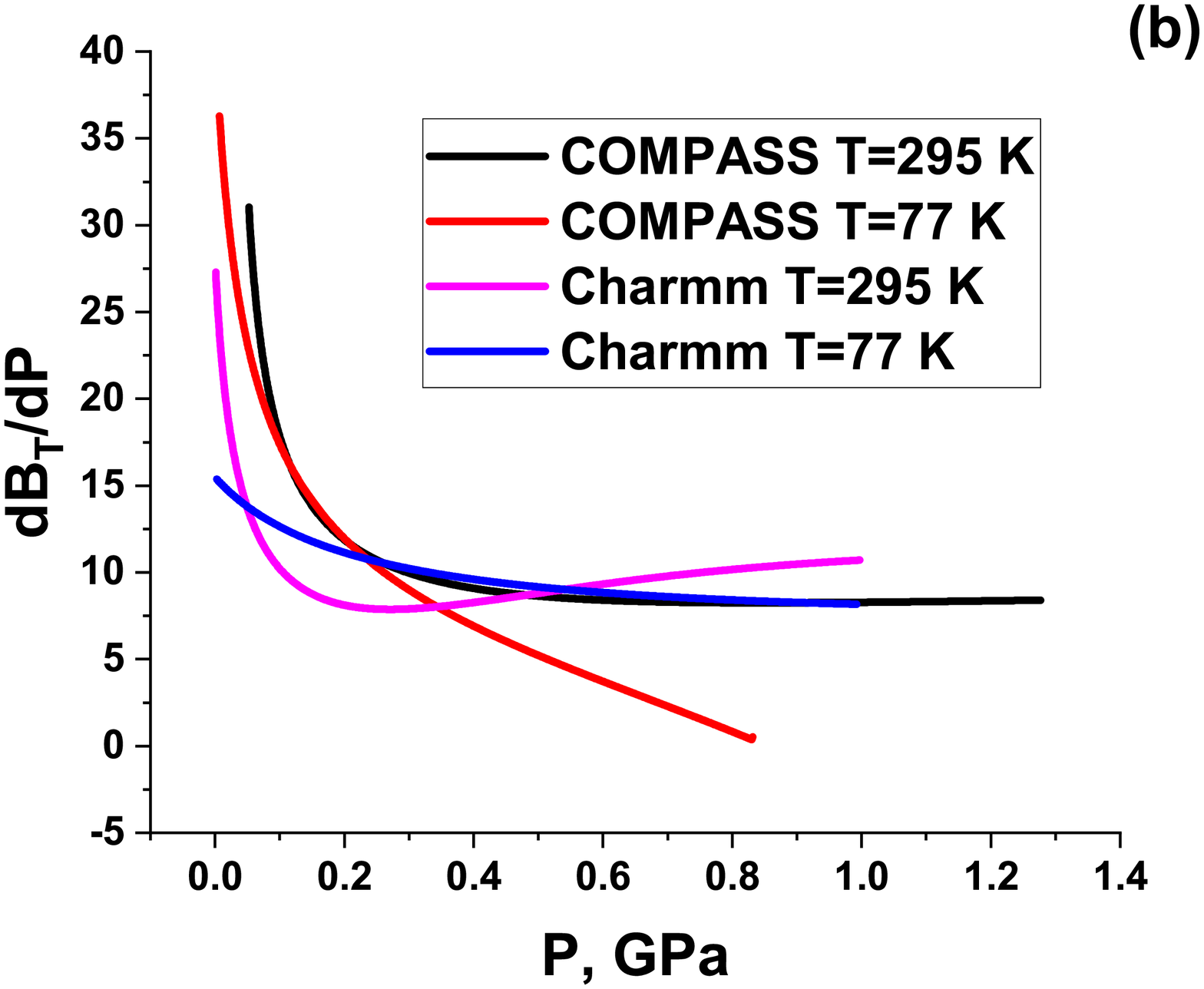}%

\caption{\label{bt-bd} (a) Comparison of experimental and calculated bulk modulus for PG at 295 and 77 K. (b) Baric derivative at 295 and 77 K
from MD with COMPASS and Charmm models.}
\end{figure}

At the same time one can see that EoS obtained by COMPASS FF at $T=295$ K has larger derivative than the one from experiment, while the slope of 
Eos obtained by Charmm
 FF is close to the experimental one. Figure \ref{bt-bd} (a) shows a comparison of bulk modulus $B_T$ obtained withing the framework
of these two force fields with the experimental data for the modulus $B_S$. Although we calculate isothermal modulus in case of molecular
simulation and adiabatic in case of experimental work, as it was shown in Supplementary materials of Ref. \cite{i2}, these moduli are 
almost indistingluishable at $T=77$ K and deviate from each other within $10 \%$ interval at room temperature. For this reason such a comparison
is possible.
One can see almost perfect agreement of the data from Charmm FF with experimental data for the pressures
above $0.2$ GPa, while COMPASS FF overestimates $B_T$ comparing to the experimental data.  

Experimental data for bulk modulus suggest that $B_T$ is almost constant withing the studied interval of pressure and takes value close to $8.0$. This
value of baric derivative corresponds to LJ system. Basing on this reasoning one can assume that the interaction between PG molecules can be roughly
approximated by the LJ potential.

The baric derivatives from molecular simulation demonstrate more complex behavior (see Fig. \ref{bt-bd} (b)).  Although baric derivative from simulation
is not constant for the case of COMPASS FF at $T=295$ K, but demonstrates assymptotic tendency to the value of $8.0$. From this reasoning one can expect that
the interaction potential of PG molecules becomes more LJ-like with pressure.

In order to check the assumption above we calculate the RDFs of centers of mass of PG molecules along $T=295$ K isotherms with COMPASS FF.
From this figure one can see that RDF of centers of mass of PG molecules cannot be described by LJ model by two reasons: 
(i) there is a pre-peak next to the first peak of $g(r)$ and (ii) the first peak of $g(r)$ demonstates a shoulder at the right-hand side of the peak.
At the same time one can see that both effects become lower as the pressure increases. Basing on this one can expect that further increase of the pressure 
may lead the center of mass RDF of PG to the shape character that the RDF of LJ system. 

\begin{figure}

\includegraphics[width=8cm, height=6cm]{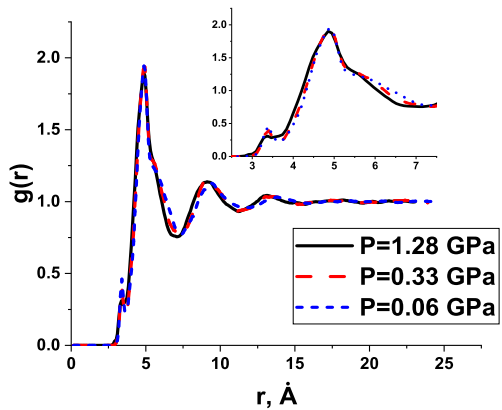}%

\caption{\label{rdf-cm} Radial distribution functions of centers of mass of PG molecules at $T=295$ K obtained with COMPASS FF.
The inset enlarges the first peak of $g(r)$.}
\end{figure}

Finally we are going to discuss the hydrogen bonding in PG. Formation of a hydrogen bond requires a polar hydrogen. In PG molecule it
corresponds to the $H_{12}$ and $H_{13}$ atoms, which are bonded to oxygens within hydroxyl groups (see Fig. \ref{pg-mol}). The hydrogens belonging
to $CH_3$ groups do not form hydrogen bonds due to their small polarity \cite{pg-ff}. 

Our calculations show that at $T=295$ K the number of hydrogen bonds does not depend on pressure: at all studied pressures the number of hydrogen
bonds is withing 410-460 in the system of 1000 molecules, i.e. less than 0.5 bonds on a molecule. Such small number of hydrogen bonds does not allow
to form a network. 

A common problem os H-bonds definition from geometrical criteria is their strong dependence on the choice of cut-off parameters. A usual approach
to selection of $r_{cut}$ is based on the first minimum of the $O-H$ partial RDF. However, there is no such a simple criterion for the angle cut-off.
In our simulations we observe that the angle cut-of strongly influence the results. For instance, if we take $\alpha_{cut}=40^o$ the number of
H-bonds doubles and becomes about 800-860 bonds per 1000 molecules of PG. However, independently on the choosen cut-offs we observe
that the number of H-bonds demonstrates very tiny dependence on pressure.

\section{Supplementary materials}

\section{Experimental methods}

Here we provide more detailed information on the experimental protocol employed in the work.

There were problems with determining the density experimentally through a change in the length of the sample, because at low pressures (up to $ 0.2$ GPa) it is difficult to 
provide all-round uniform compression, it is uniaxial in nature. Therefore, we calculated the density by integrating the equation of state. Using the definition of the isothermal bulk modulus,
$B_T=-V \left( \frac{\partial P}{\partial V} \right)_T=\rho \left( \frac{\partial P}{\partial \rho} \right)_T$, 
where $V$ is volume, we obtain that $\left( \frac{\partial \rho}{\partial P} \right)_T=\rho/B_T$   and, subsequently, for pressure dependence of density

\begin{equation}\label{eq1}
 \rho(P)=\rho_0+ \int_0^P \frac{\rho}{B_T}dP.
\end{equation}
Here one can take the atmospheric value of the initial density $\rho_0$ in the case of our pressure range. Adiabatic bulk modulus Bs can be calculated as:

\begin{equation}\label{eq2}
 B_S=\rho v_l^2 - \frac{4}{3} \rho v_t^2,
\end{equation}
where $\rho$ is the density of a substance, $v_l$  islongitudinal ultrasonic wave velocity and $v_t$ transverse ultrasonic wave velocity.
The adiabatic (isentropic) bulk modulus $B_S$ related to $B_T$ by the equation \cite{i5}

\begin{equation}\label{eq3}
  \frac{B_s}{B_T}=1+\frac{\alpha_P^2 T B_s}{\rho c_P}=1+ \xi
\end{equation}
where $\alpha_P$ is the volume thermal expansion coefficient, $c_P$ is the specific heat at constant pressure, we can calculate density from experimental ultrasonic transverse 
$v_t$ and longitudinal $v_l$ velocities as functions of pressure. From Eqs. \ref{eq1} - \ref{eq3} one can obtain direct relation between density and the experimentally measured ultrasonic velocities:

\begin{equation} \label{eq4}
  \rho(P)=\rho_0+ \int_0^P \frac{1+ \xi}{v_l^2- \frac{4}{3} v_t^2}dP.
\end{equation}

For ordinary (non-viscous) liquids Eq. \ref{eq4} simplifies to 

\begin{equation} \label{eq5}
  \rho(P)=\rho_0+ \int_0^P \frac{1+ \xi}{v_l^2}dP.
\end{equation}

In the first approximation, we can take $\xi \approx \xi_0=const$ , where $\xi_0$ can be easily calculated from the atmospheric tabulated parameters.

\section{Comparison with literature data}

\begin{figure}

\includegraphics[width=8cm, height=6cm]{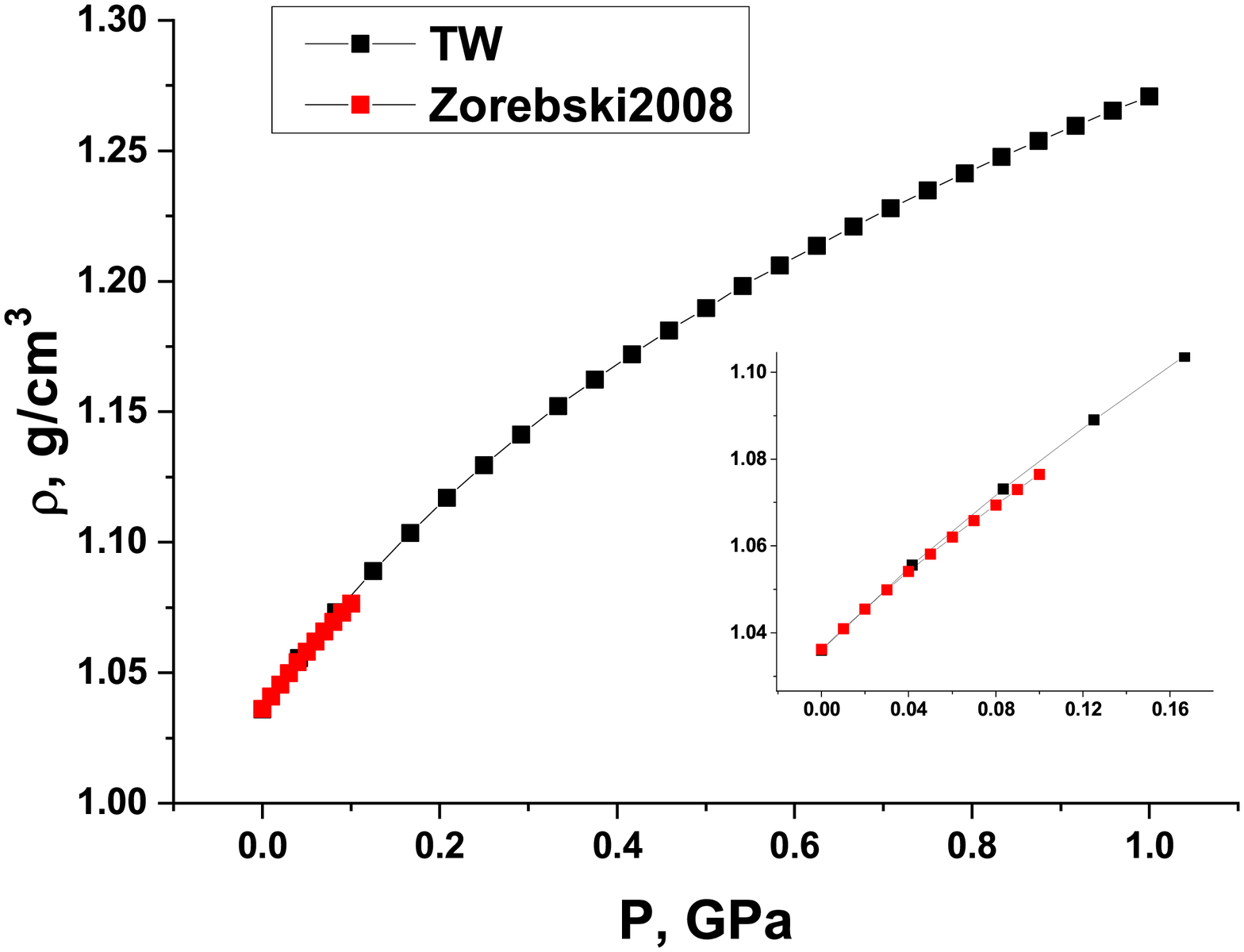}%

\caption{\label{zoreb} A comparison of our experimental data (TW) with the data from Ref. \cite{zoreb2008}. The inset
enlarges the low pressure part of the plot. Our data are at $T=295$ K, while the data from Ref. \cite{zoreb2008} at $T=293$ K.}
\end{figure}

Most of experimental works deal either with mixtures of PG with other liquids (see, for instance, Refs. \cite{m1,m2,m3,m4,m5}), or with PG oligomers 
(see, for instance, Ref. \cite{olig}). To the best of our knowledge there is just a single work \cite{zoreb2008} where
equation of state of pure PG is measured up to 101 MPa (0.101 GPa). Figure \ref{zoreb} shows a comparison of our data with the ones 
from Ref. \cite{zoreb2008}. One can see that the data are in perfect agreement: the maximum deviation of the data is about $0.3 \%$.

\section{Computational methods}

In molecular dynamics simulation we obtaine pressure as a function of density, but one of objectives of the present
work is to calculate the baric derivative of pressure, i.e. the second derivative of the data from simulation. As a 
result, even small fluctuations of the computational data may lead to great errors in the baric derivative. In order
to avoid these problems we approximate obtained equation of state (EoS) by 4-th order polynomial functions:

\begin{equation}
P=b_0+b_1 \rho +b_2 \rho^2 + b_3 \rho^3 + b_4 \rho_4,
\end{equation}
where the density $\rho$ is experssed in $g/cm^3$ and pressure $P$ in $GPa$.

The coefficients for the COMPASS at $T=295$ K: $b_0=176.16727$, $b_1=-595.92749$,
$b_2=762.16044$, $b_3=-439.70498$ and $b_5=97.34986$.

The coefficients for the COMPASS at $T=77$ K: $b_0=-45.77225$, $b_1=296.34924$,
$b_2=-543.53734$, $b_4=-97.71377$.

The coefficients for the Charmm at $T=295$ K: $b_0=243.77289$, $b_1=-941.35592$,
$b_2=1366.17016$, $b_3=-885.77148$, $b_4=217.29296$.

The coefficients for the Charmm at $T=77$ K: $b_0=197.84408$, $b_1=-657.07497$,
$b_2=828.19556$, $b_3=-475.25861$, $b_4=106.14844$.

The computational part of this work was carried out using computing resources of the federal
collective usage center "Complex for simulation and data
processing for mega-science facilities" at NRC "Kurchatov
Institute", http://ckp.nrcki.ru, and supercomputers at Joint
Supercomputer Center of the Russian Academy of Sciences (JSCC
RAS). The experimental part of the work was supported by
Russian Science Foundation (Grant 22-22-00530) and the computational part was supported by 
the Council of the President of the Russian Federation for State Support of Young Scientist (Grant MD-6103.2021.1.2).

\end{document}